\documentclass[conference]{IEEEtran}
\IEEEoverridecommandlockouts

\usepackage{cite}
\usepackage{graphicx}
\usepackage{amsmath,amssymb}
\usepackage{tabularx, booktabs}
\usepackage{color}
\usepackage{comment}
\usepackage{url}
\usepackage{bm}
\usepackage{mdframed,xcolor}
\def \am [#1]{\textcolor{red}{AM: #1}}
\def \im [#1]{\textcolor{blue}{IM: #1}}

\begin{document}
%

\title{What is the ground truth? \\ Reliability of multi-annotator data for audio tagging}

\author{\IEEEauthorblockN{Irene Mart\'{\i}n-Morat\'o, Annamaria Mesaros\thanks{This paper has received funding from Academy of Finland grant 332063 "Teaching machines to listen".}}
\IEEEauthorblockA{Computing Sciences, Tampere University\\
Tampere, FINLAND\\
Email: \{irene.martinmorato, annamaria.mesaros\}@tuni.fi
}
}


\maketitle

\begin{abstract}
Crowdsourcing has become a common approach for annotating large amounts of data. It has the advantage of harnessing a large workforce to produce large amounts of data in  a short time, but comes with the disadvantage of employing non-expert annotators with different backgrounds. This raises the problem of data reliability, in addition to the general question of how to combine the opinions of multiple annotators in order to estimate the ground truth. This paper presents a study of the annotations and annotators' reliability for audio tagging. We adapt the use of Krippendorf’s alpha and multi-annotator competence estimation (MACE) for a multi-labeled data scenario, and present how MACE can be used to estimate a candidate ground truth based on annotations from non-expert users with different levels of expertise and competence.
\end{abstract}

%

\section{Introduction}

Annotated audio is a fundamental component in training and evaluation of sound classification. Given the recent advances in environmental sound classification, including sound events classification, tagging, and detection, coupled with the use of deep learning-based solutions, availability of large datasets has become a crucial necessity. While unsupervised learning \cite{fonseca2020unsupervised} or
automatic methods for predicting labels \cite{Gemmeke2017} can provide an alternative to human-annotated data at the training stage, annotated data still plays an important role evaluation \cite{mesaros2018datasets}. 

Manual annotation of audio requires repeated listening of the given sample in order to annotate it, and relying on expert annotators makes it a slow process. For this reason, crowdsourcing has emerged as an attractive method for increasing the volume of data \cite{Cartwright2019, fonseca2019, humphrey2018openmic}. Its disadvantage is mainly that it relies on non-expert annotators, who may provide incorrect or inconsistent labels. A common processing of such multi-annotator labels in order to create the reference annotation is to aggregate them using a majority vote (consensus), as done for example in the case of OpenMIC 2018 dataset for instrument recognition \cite{humphrey2018openmic} or proposed for environmental sound classification in  \cite{Cartwright2019}.  

Still, because annotating audio requires both time and resources, having multiple annotators describe each data point remains relatively rare. In the audio domain, the CHIME-Home dataset \cite{Foster2015} was obtained using three annotators, and the final annotation was created as a majority vote. Another example is the DCASE 2013 Office Live dataset that was annotated by two persons, with both annotation sets provided with the data; within the challenge, submitted systems were evaluated against each annotator separately, and then the performance was averaged \cite{Stowell2015}.
Multiple expert annotators are more common in medical imaging for automatic diagnostic algorithms. Methods for fusing the expert annotator opinions include different strategies, from simple ones like intersection and union \cite{Kauppi2009}, to complex ones that estimate an optimal ground truth using expectation-maximization as done in STAPLE \cite{Warfield2004} or maximizing the joint agreement between annotators \cite{Kamarainen2012}. The method used to estimate the ground truth was found to have a significant effect on the evaluated performance of the system, with STAPLE causing underestimation of performance when only few annotations are available, and  consensus overestimating it \cite{Lampert2016}.

In this paper, we tackle an important research problem that has not been yet addressed in the audio domain, namely how to aggregate opinions from non-expert annotators with different backgrounds and levels of expertise in order to create a reliable ground truth for training sound classifiers. We use a subset of the publicly available TAU Urban Acoustic Scenes 2019 dataset \cite{Mesaros2019_DCASE} that we annotate using sound event tags. 
We estimate annotators' competence  and inter-annotator agreement using established statistical tools, and compare different aggregation procedures for creating the reference annotation. We show that a low agreement does not necessarily reflect a low annotator reliability, instead it partly reflects the difficulty of the annotation task. 

The paper is organized as follows: Section \ref{sec:methods} introduces the annotator competence and agreement measures we will use in our analysis, Section \ref{sec:data_collection} presents the data we use and the annotation process, Section \ref{sec:data_analysis} presents the analysis of the collected data and further experiments. Finally, Section \ref{sec:concl} presents conclusions and future work.

\section{Annotator and annotation analysis}
\label{sec:methods}

We propose to adapt and employ a collection of methods that are more familiar to those working in computational linguistics. 
\textit{Labeling} is the process of assigning a label to an item by an \textit{annotator}. In our study, we deal with \textit{multi-label annotation}, i.e. an item (in our case an audio file) is assigned one or multiple labels from a pre-defined set of labels. 

The largest datasets for audio classification typically rely on user-generated material available as web audio, for which labels can be inferred from user-generated data. For example AudioSet \cite{Gemmeke2017} consists of automatically labeled and partially verified audio, but has an estimated label error of above 50\% for 30\% of its classes\footnote{See https://research.google.com/audioset/dataset/index.html for an explanation of the quality assessment. Information accessed January 2021}; FSDnoisy18k \cite{fonseca2019} was crowdsourced and a subset of it was curated by experts. These are two examples showing the trade-off between accepting noisy labels and the effort necessary for curation. The reliability of the annotation process for audio data and its outcome, rarely analyzed before, is something we aim to do in this study. 

\subsection{Annotator competence estimation}

When a large pool of annotators that annotate partially the same data is available, the competence of these annotators can be estimated using MACE - Multi-Annotator Competence Estimation \cite{hovy2013}. The method allows identification of trustworthy annotators and prediction of correct underlying labels, by using an unsupervised model that learns from redundant annotations. 

The model considers that annotator $j$ produces label $A_{ij}$ on instance $i$. The annotated label depends on the true label $T_i$, and whether annotator $j$ is spamming. Annotation behavior is modeled by binary variable $S_{ij}$ drawn from a Bernoulli distribution with parameter $(1 -\theta_j)$. The behavior assumes that when an annotator is not spamming on instance $i$ ($S_{ij} = 0$), the annotation $A_{ij}$ corresponds to the true label. When the annotator is spamming, $S_{ij} = 1$, $A_{ij}$ is sampled from a multinomial distribution with parameter vector $\xi_j$. The annotations $A_{ij}$ are observed, the true labels $T_i$ and the spamming indicators $S_{ij}$ are unobserved. The model parameter $\theta_j$ specifies the probability of trustworthiness for annotator $j$, while $\xi_j$ determines the spamming behavior of annotator $j$.

The model parameters are estimated using the expectation maximization algorithm, to maximize the probability of the observed data \cite{hovy2013}: 
\begin{equation}
    P(\mathbf{A}; \theta,\xi) = \sum_{T,S} \left[ \prod_{i=1}^N P(T_i)  \prod_{j=1}^M P(S_{ij}; \theta_j) P(A_{ij}|S_{ij}, T_i;\xi_i) \right]
\end{equation}
where $\mathbf{A}$ is the matrix of annotations, $\mathbf{S}$ is the matrix of competence indicators, and $\mathbf{T}$ is the vector of true labels.
The method was shown to produce predicted labels very accurately in comparison with ground truth data on a few tasks. At the same time, the model's $\theta_j$ was shown to correlate strongly with annotator proficiency \cite{hovy2013}. 

We use MACE to study the behavior of our annotators and to predict different sets of aggregated labels. It is important to note that MACE does not discard annotators, but weighs their opinion based on their competence, which results in a different procedure than majority voting which trusts and weighs all annotators equally. 

\subsection{Inter-annotator agreement}

Many measures that assess inter-annotator agreement are developed for only two annotators. In addition, simple measures like percentage of agreement or correlation suffer from various biases related to chance agreement and statistical independence of annotators and annotated data \cite{Krippendorff2011}. 
We select for our analysis Krippendorff's alpha as a general agreement metric that is able to cope with more than two annotators per item and with missing data (overlap in annotated items only among few of the annotators).

Krippendorff's alpha is defined as: 
 \begin{equation}
  \alpha = 1 -  \frac{ D_{o} }{D_{e}} 
 \end{equation}
 where 
$D_{o}$ is the observed disagreement and $D_{e}$ is the expected disagreement. In the case of multiple annotators $m$, multiple nominal categories, and missing values, the formulation uses \textit{nominal} $\alpha$ defined as \cite[pp.230-231]{Krippendorff2004}:

 \begin{equation}
_{nominal}\alpha = 1 - (n - 1) \frac{n- \sum\nolimits_{c}o_{cc} }{ n^2 - \sum\nolimits_{c}n_{c}^2 } 
 \end{equation}
where:
 \begin{equation}
  o_{ck} = \sum\nolimits_{u} \frac{\text{Number of $c$-$k$ pairs in $u$}}{m_u - 1}
 \end{equation}
is the number of observed coincidences of two categories $c$ and $k$ assigned to the same item by two different annotators, and $m_u$ is the number of values assigned to item $u$ (number of annotators that labeled this item). Observed coincidences are calculated based on a coincidence matrix that considers the annotators interchangeable, therefore pairing the contingencies in both directions: if $x_{ck}$ is the number of times a particular observer uses $c$ while the other uses $k$, then the number of coincidences is $o_{ck} = x_{ck} + x_{kc}$. 

Krippendorff's alpha is reported in \cite{Cartwright2019} for annotation of audio, but no discussion on its values for the annotated data is provided.  Krippendorff’s alpha has been also used to measure inter-annotator agreement for images \cite{nassar2019} and video annotations \cite{Park2012}.

\section{Multi-annotator data collection}
\label{sec:data_collection}

The dataset used in our experiments is a subset of TAU Urban Acoustic Scenes 2019 \cite{Mesaros2019_DCASE}, consisting of audio from three acoustic scenes (airport, public square, and park). The audio files are 10 seconds long, and some of them are consecutive segments of one long recording from a single location. 
Each audio file was annotated by five different annotators, following a single-pass multi-label annotation procedure \cite{Cartwright2019}, in which the annotator selected for one audio file a number of labels presented as a list. Candidate labels were \textit{birds singing, dog barking, adults talking, children voices, traffic noise, music, footsteps, siren, announcement speech} and \textit{announcement jingle}. According to the annotation procedure, a positive label is explicitly provided by the annotator, while a negative label is implicit, by not being selected. 

The annotation platform was a simple web-based interface that presented users with audio files to annotate one by one. Annotators registered using their email, which permitted the annotation process to be paused and resumed later on next login. A set of instructions and examples of annotation were provided at the beginning. Annotators were instructed to work in small batches and to use good quality headphones. It was allowed to listen to each audio file multiple times before selecting one or more of the candidate labels.
A total of 133 annotators, students taking an audio signal processing course, were randomly assigned a maximum of 131 files to annotate. The total number of files annotated is 3930. Annotators were assigned into 30 groups, aiming that  each group will provide annotations to the same set of files.

Because neither MACE nor the inter-annotator agreement metrics are defined for multi-labeled items, we represent the annotations as a set of binary \textit{yes/no} labels per file, with explicit/implicit presence as explained before. In consequence, each (file, label) pair is considered an independently annotated item, equivalent to a multiple-pass binary annotation \cite{Cartwright2019}. However, because the annotation process did not request producing the labels themselves, we consider that this assumption has sufficient grounds. 

Complete annotations are represented as a matrix containing the answers of all annotators, illustrated in Table \ref{tab:annotations}. Each row refers to a (file, label) item, and each column represents the answer of one annotator in the format $[0,1,-]$, marking the presence (1, explicit) or absence (0, implicit) of this label within the audio file; "$-$" indicates that this file was not assigned to this specific annotator.
The resulting matrix contains a total of 39300 items.

\section{Data analysis}
\label{sec:data_analysis}

\begin{figure}
    \centering
    \includegraphics[width=1\columnwidth]{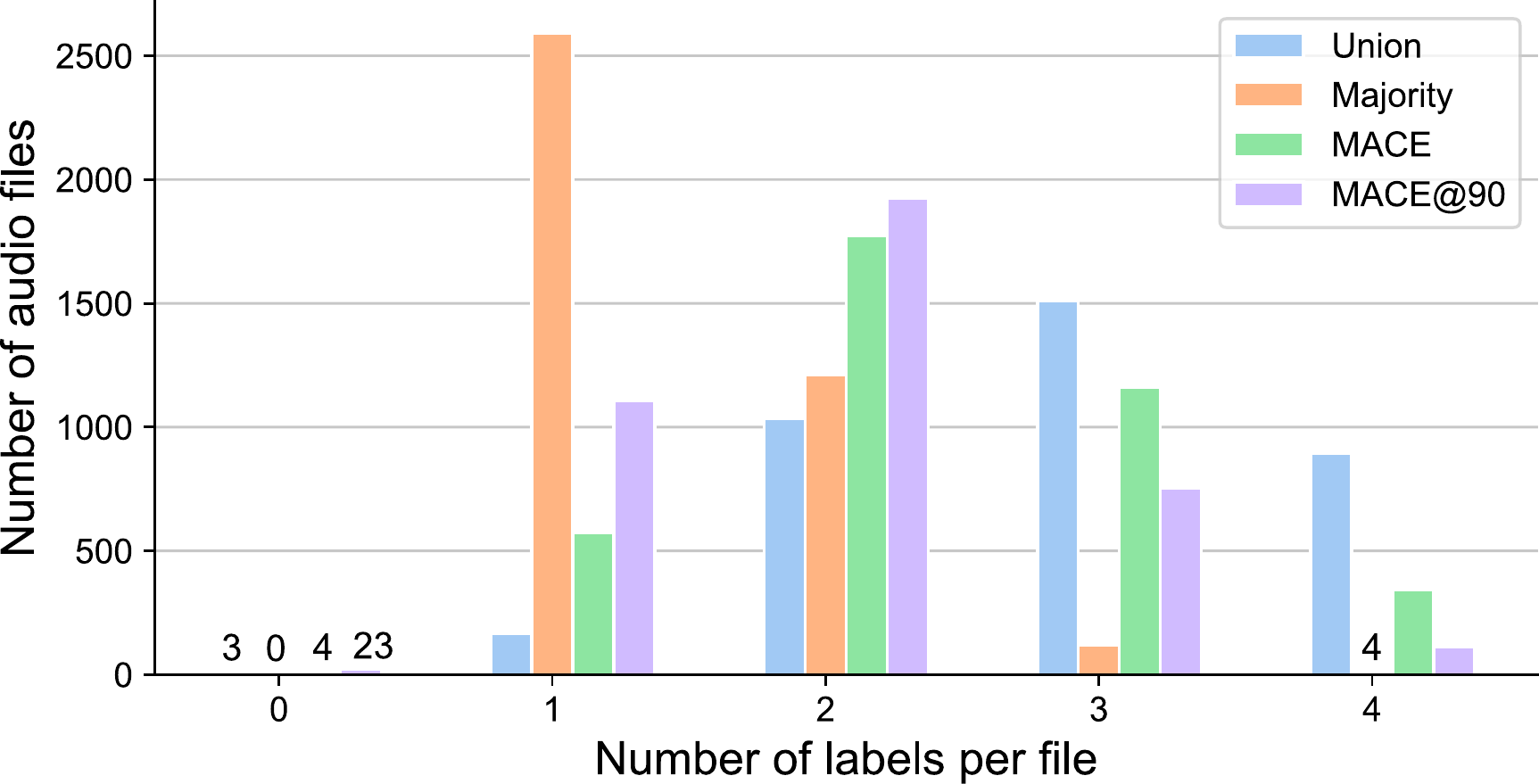}
    \caption{Average number of labels per file produced by different approaches to estimate the true labels: union, majority vote, MACE, and MACE@90}
    \label{fig:labels-per-file}
    \vspace{-14pt}
\end{figure}

We first consider the aggregation of multiple annotations. The simplest one, union, assigns a label to a file if at least one of the annotators has considered it active. The most commonly used aggregation method, majority vote, assigns a label to a file if most annotators have considered it active. 
The statistics of the resulting classes are presented in Table \ref{tab:label-stats} for the individual classes (first two columns), while Fig. \ref{fig:labels-per-file} shows the resulting number of labels per file.  
The resulting annotations are largely unbalanced, with the most common label \textit{adults talking} being assigned to 3168 files, and least common \textit{announcement jingle} to 116 files. Majority voting reduces their frequency in the resulting annotation to 2401 and 8, respectively.

\begin{table}[]
    \centering
    \caption{Annotation matrix example with explicit/implicit annotations produced by $m$ annotators}
\begin{tabular}{r|c|c|c|c|c|c}
items (file, label) & 1 & 2 & 3 & 4 & .. & m \\
\midrule
     airport-Paris-0, footsteps & 1 & 1 & 0 & 1 & ... & -  \\
     airport-Paris-0, adults-talking & 0 & 0 & 0 & 1 & ... & -  \\
     airport-Paris-0, dog-barking & - & - & - & - & ... & 1  \\
     ... & ...  & ... & ... & ...  & ... &    \\
     airport-Helsinki-4, footsteps & 1 & 1 & 1 & - & ... & -   \\
\bottomrule
\end{tabular}
\label{tab:annotations}
\vspace{-10pt}
\end{table}

\begin{table}[]
    \centering
    \caption{Statistics of class labels resulting from different methods of combining the multiple annotations}
\begin{tabular}{r|c|c|c|c}
\toprule
class labels & union & maj. vote & MACE & MACE@90  \\
\midrule
adults talking & 3168 & 2401 & 2983 & 2831\\
footsteps & 2560 & 859 & 1969 & 1583\\
traffic noise & 2418 & 680 & 1713 & 1178\\
children voices & 1467 & 513 & 1046 & 821\\
birds singing & 1332 & 672 & 1035 & 855\\
music & 306 & 106 & 212 & 174\\
ann. speech & 273 & 73 & 148 & 108\\
dog barking & 177 & 42 & 108 & 79\\
siren & 177 & 38 & 99 & 61\\
ann. jingle & 116 & 8 & 38 & 16\\
\bottomrule
\end{tabular}
\label{tab:label-stats}
\end{table}

\subsection{Predicting ground truth with MACE}

As explained, MACE predicts the true labels by estimating the annotators' competences and their effect on the true labels within the same model. This creates a weighing procedure on the different opinions which is dependent on how trusted the respective annotator is. In addition, the produced estimation for ground truth can be constrained using a threshold $n$, with MACE@n containing the $n$\% of predicted labels for which the method is most confident \cite{hovy2013}. 

The resulting statistics of the estimated ground truth in terms of number of produced labels and number of labels per file are presented in Table \ref{tab:label-stats} and Fig. \ref{fig:labels-per-file} for comparison with the union and majority vote. With MACE, the number of labels estimated for the ground truth is significantly higher than using majority vote for all categories, indicating that for some cases a minority of annotators is reliable enough to justify the label. Even when eliminating the least confident 10\% of predictions (MACE@90), the number of resulting labels is higher than with the majority vote, showing that this method has the potential to overcome the problem of missing labels caused by an insufficient number of votes, which can cause label noise for learning \cite{fonseca2020}. 

\subsection{Annotator competence analysis}

\begin{figure}
    \centering
    \includegraphics[width=1\columnwidth]{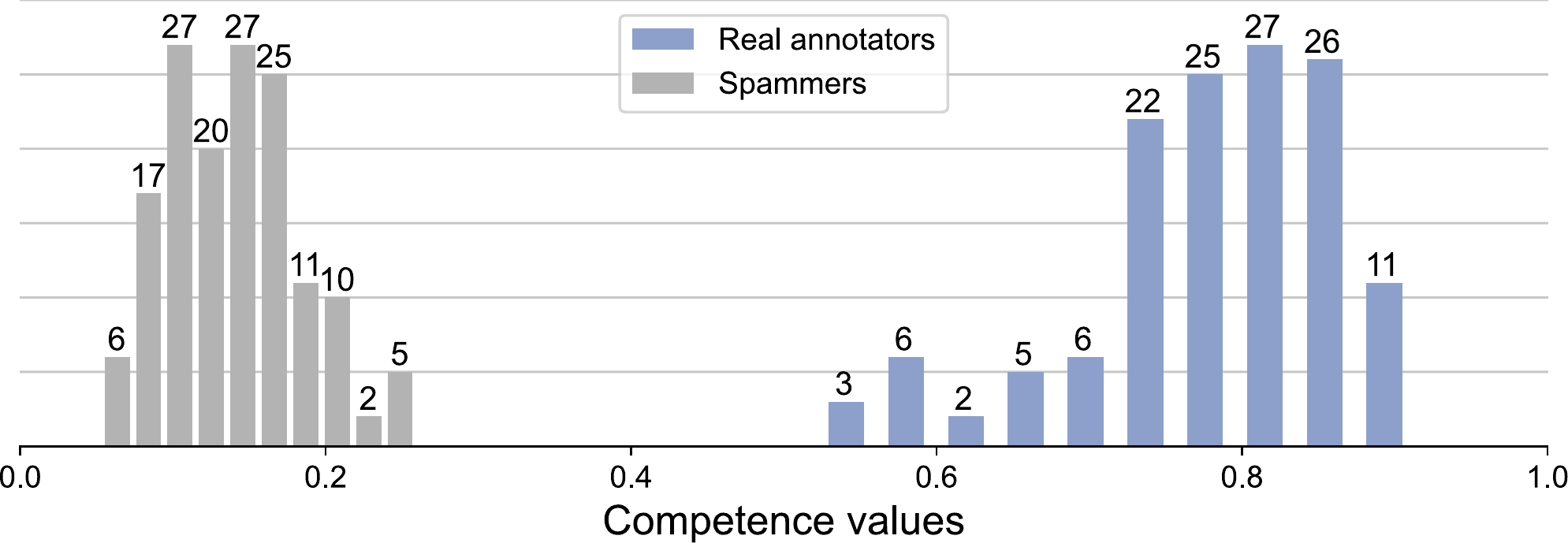}
    \caption{Annotator competence estimated using MACE}
    \label{fig:mace-competence}
    \vspace{-14pt}
\end{figure}

The estimated competence of our annotators, obtained using MACE, is illustrated in Fig. \ref{fig:mace-competence}. We observe that there are a number of annotators that are highly trustable (64 over 0.8), while a small number of them have much lower estimated competence. Even though the annotators do seem mostly reliable, the agreement on the labels is not very high, with Krippendorff's alpha for the entire dataset being 0.696.

We hypothesize that inter-annotator disagreement can come from two sources. One is the annotator competence: an annotator who does not pay attention to the task and completes it at random will not have high agreement with an annotator who is very diligent about the task. 
A second source of disagreement is the annotator's personal experience and perception. Experiments in cognitive psychology have shown that life experience brings much subjectivity in categorization \cite{guastavino2018everyday}. The audio data in our experiment is recorded in the wild, with no control over the sound sources present, their prominence in the scene and their overlaps, which makes it rather difficult to annotate and allows personal interpretation. In addition, some studies have shown that visual stimuli help with audio annotation \cite{cartwright2017}, but our experiment did not provide any visual information.

In absence of the gold standard (which would allow us to estimate the upper bound for agreement and evaluate the estimated ground truth), we simulate the lower bound. 
We simulate a group of spammer annotators for the task, that provide \textit{yes/no} indicators per file for the set of 10 labels. We create 150 random annotators, with each being randomly assigned a number of 130 files from the set of 3930 available. We then analyze their output in terms of labels statistics, majority vote, MACE competence, and inter-annotator agreement metrics. 
The  estimated competence of these spammers, presented in Fig.\ref{fig:mace-competence}, shows that even though the real annotators disagree on the labels, they are in fact diligent and not answering at random. The distribution of labels per file for the random annotators is much more uniform, and inter-annotator agreement (Krippendorff's alpha) is practically 0. This suggests perception differences as being the main cause of disagreement of annotators, but because we cannot separate the effects of the two in the data, we cannot draw a definite conclusion.

\subsection{Inter-annotator agreement and improving data reliability}

Krippendorff's alpha was calculated for the overall data, and separately for each class. The results are  presented in Table \ref{tab:alpha-values}. We observe a wide variation in the class-wise agreement, with the highest agreement on the more rare \textit{dog barking} class. On the other hand, the more frequent classes \textit{footsteps} and \textit{traffic noise} have similar frequency in our data but very different agreement values. Their different acoustic characteristics also indicate perception as a reason for disagreement, as explained in the previous paragraph. 

A straightforward way to improve the data reliability when we have knowledge about annotators competence is to eliminate annotations produced by the least trusted annotators, in order to obtain a set of annotations which is produced by the most reliable ones. Of course, gradual elimination of annotators will result in a reduced set of annotations available, until the extreme case of having only a single annotator. Table \ref{tab:alpha-values} presents the calculated alpha for the case of only using annotators with estimated competence of at least 0.6 (124 annotators) and at least 0.8 (64 annotators). 

\begin{table}[]
    \centering
    \caption{Krippendorff's alpha for selected subsets of annotations}
\begin{tabular}{r|c|c|c}
\toprule
class-wise & all annot & competence$>0.6$ & competence$>0.8$\\
\midrule
adults talking & 0.676 & 0.690 & 0.717\\
footsteps & 0.271 & 0.284 & 0.236  \\
traffic noise & 0.590 & 0.607 & 0.635\\
children voices & 0.712 & 0.714 & 0.729 \\
birds singing & 0.613 & 0.619 & 0.657 \\
music & 0.606 & 0.615 & 0.679 \\
ann. speech & 0.485 & 0.501 & 0.548 \\
dog barking & 0.713 & 0.730 & 0.764  \\
siren & 0.550 & 0.569 & 0.624 \\
ann. jingle & 0.404 & 0.430 & 0.525 \\ 
\midrule
\textbf{overall} & \textbf{0.696} & \textbf{0.708} & \textbf{0.745} \\
\bottomrule
\end{tabular}
\label{tab:alpha-values}
\end{table}

\begin{figure}
    \centering
    \includegraphics[width=0.8\columnwidth]{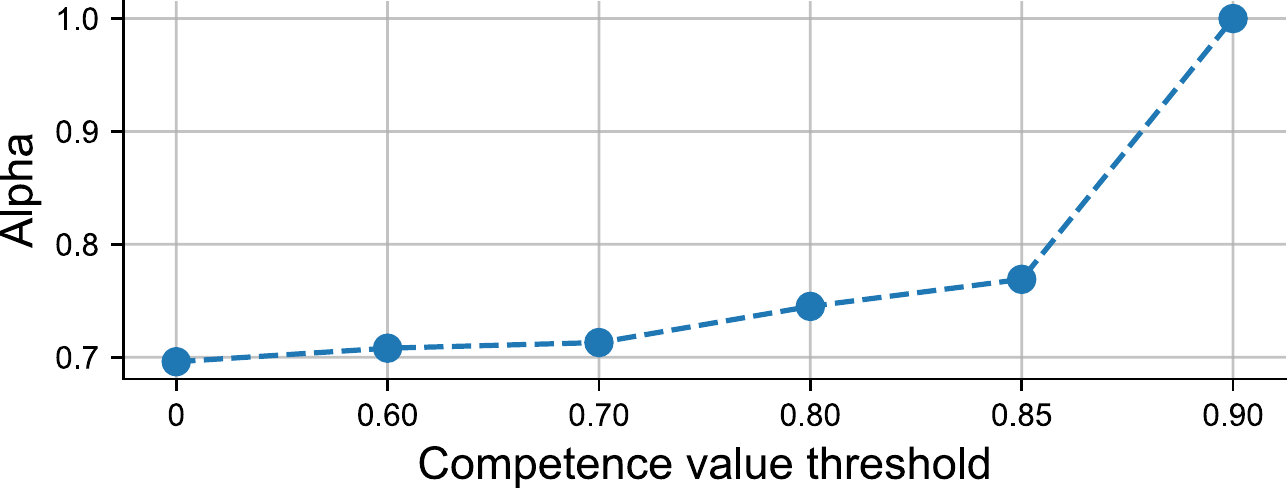}
    \caption{Alpha values when gradually removing annotators with competence values under a given threshold, until only one is left}
    \label{fig:alpha-values}
    \vspace{-14pt}
\end{figure}

All agreement values increase when using the more reliable annotators, except for \textit{footsteps}, while the overall agreement increases significantly when using the top annotators. Figure \ref{fig:alpha-values} shows the evolution of $\alpha$ when annotators under a given competence are gradually eliminated, with the final case being a single annotator.
As a comparison, we note that standards adopted in social sciences consider a 0.8 agreement reasonable, and consider values between 0.667 and 0.8 only for drawing tentative conclusions. However, Krippendorff argues that the acceptable level of agreement must be chosen depending on the costs of drawing invalid conclusions \cite[p. 241]{Krippendorff2004}. Therefore we can state that the employed methods allow creating reference annotations that can be trusted for training and evaluation of acoustic models, compared to noisy data\footnote{The produced raw annotations and resulting aggregated ground truth will be published as open data on paper acceptance. A link to the data will be added in the camera ready paper.}.

\section{Conclusions and future work}
\label{sec:concl}

This paper presented a study of annotator and annotations reliability for crowdsourced audio tags. We showed that the aggregation of raw multi-annotator labels using annotator competence estimation produces a plausible and trustable ground truth, with gradually improving levels of agreement in the data. However, in our experiment we cannot evaluate the correctness of the estimated ground truth. For this reason, we plan to repeat this experiment in controlled conditions, using generated synthetic data for which ground truth is produced at the same time with the audio. We will try to mimic as closely as possible the classes and acoustic characteristics of the data used in the presented experiment.





%


\bibliographystyle{IEEEtran}
\bibliography{refs}

\end{document}